\documentclass[twocolumn,showpacs,preprintnumbers,superscriptaddress]{revtex4}

\usepackage{amssymb}
\usepackage{graphicx}
\usepackage{epsfig}
\usepackage{amsmath,amssymb}

\begin{document}

\title{ Simulating spin-charge separation with light}
\date{\today }
\author{Dimitris G. Angelakis}
\email{dimitris.angelakis@gmail.org}
\affiliation{Science Department, Technical University of Crete, Chania, Crete, Greece,
73100}
\affiliation{Centre for Quantum Technologies, National University of Singapore, 2 Science
Drive 3, Singapore 117542}
\author{Mingxia Huo}
\affiliation{Centre for Quantum Technologies, National University of Singapore, 2 Science
Drive 3, Singapore 117542}
\author{Elica Kyoseva}
\affiliation{Centre for Quantum Technologies, National University of Singapore, 2 Science
Drive 3, Singapore 117542}
\author{Leong Chuan Kwek}
\affiliation{Centre for Quantum Technologies, National University of Singapore, 2 Science
Drive 3, Singapore 117542}
\affiliation{National Institute of Education and Institute of Advanced Studies, Nanyang
Technological University, 1 Nanyang Walk, Singapore 637616}

\begin{abstract}
 In this work we show  that stationary light-matter excitations generated inside a hollow one-dimensional  waveguide filled with  atoms, can be made to generate a photonic two-component Lieb Liniger model.  We explain  how to prepare and drive the atomic system to a strongly interacting regime where spin-charge separation could be possible. We then proceed by explaining how to measure the corresponding effective spin and charge densities and velocities through standard optical methods based in measuring dynamically the emitted photon intensities or by analyzing the photon spectrum. The relevant interactions exhibit the necessary tunability  both to generate and efficiently observe spin charge separation with current technology.
\end{abstract}
\maketitle


One of the most counterintuitive characteristics of one dimensional electron gases is spin-charge separation.
In this case the electrons cease to behave as single particles comprised of spin and charge. Instead collective excitations appear carrying only charge (and no spin) or only spin (and no charge) which can propagate through the system with different velocities\cite{Girardeau}. Spin charge separation was predicted in tunneling experiments in metallic chains \cite{metallic}, organic conductors \cite{organic}, carbon nanotubes \cite{SCC3} and more recently in quantum wires \cite{SCC4}.
However due to the complexity of the structures used, measuring the spectral function and observing distinct spinon and holon branches has yet to be conclusive. In parallel with these works, artificially engineered many-body systems that could simulate condensed matter effects in well controllable environments have been developed the last decade. Cold atoms and ions traps are the most famous example \cite{Bloch,Jaksch,ions}, and strongly interacting photons (SIP) the most recent developments. Proposals to observe spin charge separation have also been in place for some time in cold atoms, including both bosonic and fermionic species \cite{Recati,Paredes03,Paredes04,Kecke,Kollath,Kleine}. However the lack of necessary  individual accessibility and measurement, and the challenges in trapping and cooling especially fermionic gases make current results inconclusive so far

SIPs on the other hand, as hybrid light-matter quantum simulators promise to provide the necessary extra manipulation and measurement lacking so far from other proposals. The efficient quantum simulation of photonic and polaritonic Mott transitions and the crystallization of photons was shown to be possible using both arrays of coupled resonators or stationary polaritons in atomic gases \cite{angelakis07,hartmann06,greentree06,Chang,Aichhorn,coherent control of photon emission}. We show here for the first time that spin-charge separation could be efficiently observed in a such strongly correlated quantum optical system. 

In parallel with the seminal works in Luttinger liquid and in cold atoms physics, significant progress has also been made in a different field. Quantum nonlinear optics has shown that strong interactions between light pulses, even at the single photon level is possible with numerous applications in photon switching and quantum memories. Employing Electromagnetically Induced Transparency techniques (EIT), few photons pulses can be slowed down, stored in atomic medium and then efficiently recovered again in an all coherent manner \cite{DP1,DP2,DP3}. In this work we show  that stationary light-matter excitations generated inside a hollow one-dimensional  waveguide filled with atoms, can be made to generate a photonic two-component Lieb Liniger model. Moreover  the relevant interactions exhibit the necessary tunability  both to generate and efficiently observe spin charge separation.

\begin{figure}
\centering
\includegraphics[scale=0.4]{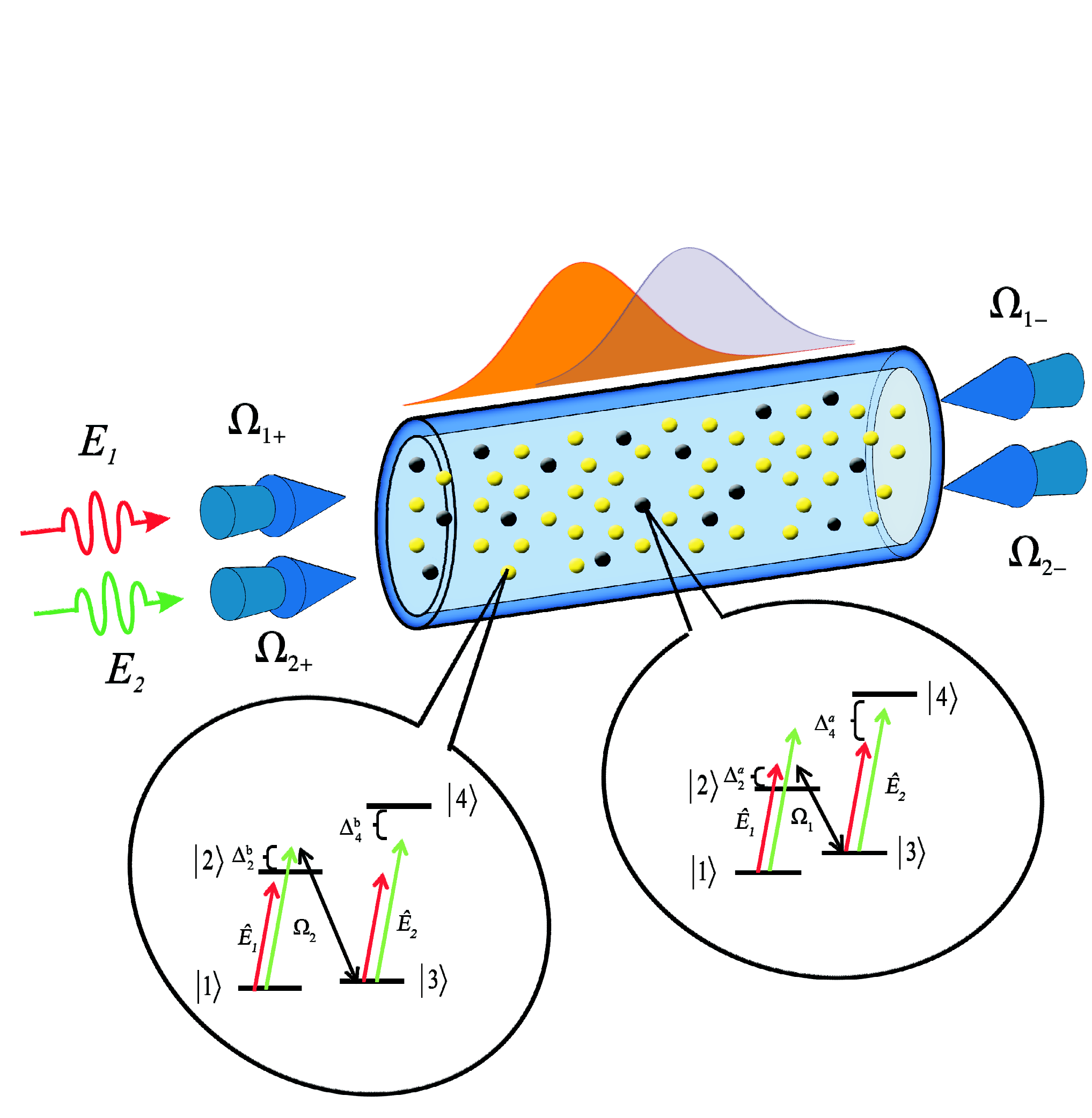}
\caption{A schematic of  system under consideration. A hollow fiber with two quantum light fields $E_{1,+}, E_{2,+}$,  and two pairs of classical fields $\Omega _{1,\pm}$ and $\Omega _{2,\pm}$  propagating towards the left and right directions. The fiber is filled with two atomic gases a and b. Initially two coherent few photon optical pulses carried by  $E_{1,+}, E_{2,+}$  are incident from one direction, say the left side. By adiabatically switching  off the corresponding control fields  the quantum pulses  can be slowed down and eventually mapped to stationary light-matter excitation (polaritons). Appropriate tuning of the couplings of the light fields to the corresponding atomic
 transitions, forces the trapped polaritons to  behave as 1D quantum liquid of two types of bosons obeying the Lieb Liniger model\cite{Girardeau}. The available tunability of the effective interaction parameters allows us to reach the spin charge separation regime. Subsequent release of the trapped polaritons through their coherent mapping back to propagating light pulses, allows for the efficient measurement of both the dynamics of the propagation of the spin and charge quasiparticles or of the spectral function characteristic of the effect taking place. This is done though standard quantum optical measurement techniques measuring correlations in the intensities of the output fields
 and by analyzing the optical spectrum}
\label{fig1}
\end{figure}

Consider a  waveguide, a hollow fiber for example filled with two types of atomic gases a and b. Two quantum light fields and two classical fields $E_{1,\pm}, E_{2,\pm}$  $\Omega _{1,\pm}$ and $\Omega _{2,\pm}$ respectively  can propagate towards the left and right directions and couple to the atoms as shown in Fig. 1. Initially two resonant optical pulses carried by  $E_{1,+}, E_{2,+}$  are incident from one direction, say the left side. They are injected into the waveguide with the co-propagating control fields $\Omega _{1,+}$ and $\Omega _{2,+}$ initially turned on. The Hamiltonian in the Schr\"{o}dinger picture can be expressed as a sum of two independent parts describing the evolution of the different atomic species $\mathsf{H}=\mathsf{H}^{\textnormal{a}} + \mathsf{H}^{\textnormal{b}}$ with
\begin{eqnarray}
\mathsf{H}^{\textnormal{x}} = &-& \hbar n^{\textnormal{x}}_{z} \sum_{i=1}^{2}
\int dz \{-\omega^{\textnormal{x}}_{33}\sigma^{\textnormal{x}}_{33} + (-\omega_{\textnormal{q}}^{(i)}+\Delta^{\textnormal{x}}_{2}) \sigma^{\textnormal{x}}_{22} \notag \\
&+& (-\omega^{\textnormal{x}}_{33} - \omega_{\textnormal{q}}^{(i)} - \Delta^{\textnormal{x}}_{4}) \sigma^{\textnormal{x}}_{44} + g_{i}^{\textnormal{x}} \sqrt{2\pi} (\sigma^{\textnormal{x}}_{21} + \sigma^{\textnormal{x}}_{43}) \times \notag \\
&& \left( \hat{E}_{i,+} \mathrm{e}^{\mathrm{i} (k_{\textnormal{qu}}^{(i)}z - \omega_{\textnormal{qu}}^{(i)}t)} + \hat{E}_{i,-} \mathrm{e}^{\mathrm{i} (- k_{\textnormal{qu}}^{(i)}z - \omega _{\textnormal{qu}}^{(i)}t)}\right) \notag \\
&+& \left( \Omega_{i,+}(t) \mathrm{e}^{\mathrm{i}(k_{\textnormal{cl}}^{(i)}z - \omega^{(i)}_{\textnormal{cl}}t)} + \Omega_{i,-}(t) \mathrm{e}^{-\mathrm{i}(k_{\textnormal{cl}}^{(i)}z + \omega^{(i)}_{\textnormal{cl}}t)}\right) \sigma^{\textnormal{x}}_{23} \notag \\
&+& \mathsf{H.c.} \}
\end{eqnarray}

with ${\textnormal{x}}=\{\textnormal{a}, \textnormal{b} \}$ labeling the two atomic species a and b. Here, the continuous collective atomic spin operators $\sigma^{\textnormal{x}}_{pq} =\sigma^{\textnormal{x}}_{pq}\left( z,t\right)$ $(p,q = \{1,...,4\})$ describe the average of $\left\vert p\right\rangle^{\textnormal{x}} \left\langle q\right\vert^{\textnormal{x}}$ over the atoms of type x in a small but macroscopic region around spacial coordinate $z$. The densities of the two species atoms in the same region are assumed to be different and equal to $n^{\textnormal{a}}_{z}$ and $n^{\textnormal{b}}_{z}$ respectively. Moreover $\hat{E}_{i,\pm}(z,t)$ and $\Omega_{i,\pm}(t)$, with $i=\{1,2\}$, are the two quantum and two classical fields slowly varying operators with frequencies $\omega^{(i)}_{\textnormal{qu}}$ and $\omega^{(i)}_{\textnormal{cl}}$, and wave vectors $k_{\textnormal{qu}}^{(i)}$ and $k_{\textnormal{cl}}^{(i)}$ respectively. Both quantum fields $\hat{E}_{1,\pm }$ and $\hat{E}_{2,\pm }$ drive four possible atomic transitions and for simplicity we assume that each field couples with the same strength to the two transitions of the different atoms. The field $\hat{E}_{1,\pm }(z,t)$ is detuned by $\Delta_{2}^{\textnormal{a}}$ from the transition $|1\rangle^{\textnormal{a}} \rightarrow |2\rangle^{\textnormal{a}}$ and by $\Delta_{4}^{\textnormal{a}}$ from $|3\rangle^{\textnormal{a}} \rightarrow |4\rangle^{\textnormal{a}}$. Similarly, the quantum field $\hat{E}_{2,\pm }(z,t)$ is off-resonant from $|1\rangle^{\textnormal{b}} \rightarrow |2\rangle^{\textnormal{b}}$ and $|3\rangle^{\textnormal{b}} \rightarrow |4\rangle^{\textnormal{b}}$ by $\Delta_{2}^{\textnormal{b}}$ and $\Delta_{4}^{\textnormal{b}}$ as shown on Fig. \ref{fig1}. Finally, the applied classical control beams with Rabi frequency $\Omega_{1,\pm}(t)$ and $\Omega_{2,\pm}(t)$ couple to both atoms and drive the transitions $|3\rangle^{\textnormal{x}} \rightarrow |2\rangle^{\textnormal{x}}$.
The evolution of the slowly-varying quantum operators $\hat{E}_{i,\pm}(z,t)$ is given by four Maxwell-Bloch equations
\begin{equation}
\left( \frac{\partial}{\partial t} \pm \nu^{(1)} \frac{\partial}{\partial z} \right)
\hat{E}_{1,\pm}(z,t) = \mathrm{i} \sqrt{2\pi } n^{\textnormal{a}}_{z} g^{\textnormal{a}}_{1}\left(
\sigma^{\textnormal{a}}_{12,\pm}(z,t) + \sigma^{\textnormal{a}}_{34,\pm}(z,t)\right),
\label{E1}
\end{equation}
and
\begin{equation}
\left( \frac{\partial}{\partial t} \pm \nu^{(2)} \frac{\partial}{\partial z} \right)
\hat{E}_{2,\pm}(z,t) = \mathrm{i} \sqrt{2\pi } n^{\textnormal{b}}_{z} g^{\textnormal{b}}_{2}\left(
\sigma^{\textnormal{b}}_{12,\pm}(z,t) + \sigma^{\textnormal{b}}_{34,\pm}(z,t)\right),
\label{E2}
\end{equation}
where we have introduced the slowly varying collective operators $\sigma^{\textnormal{a}}_{pq}=\sigma^{\textnormal{a}}_{pq,+}(z,t) \mathrm{e}^{\mathrm{i}k^{(1)}_{\textnormal{qu}}z} + \sigma^{\textnormal{a}}_{pq,-}(z,t) \mathrm{e}^{-\mathrm{i}k^{(1)}_{\textnormal{qu}}z}$ and $\sigma^{\textnormal{b}}_{pq} = \sigma^{\textnormal{b}}_{pq,+}(z,t) \mathrm{e}^{\mathrm{i}k^{(2)}_{\textnormal{qu}}z} + \sigma^{\textnormal{b}}_{pq,-}(z,t) \mathrm{e}^{-\mathrm{i}k^{(2)}_{\textnormal{qu}}z}$. In the derivation of Eqs. (\ref{E1},\ref{E2}) we have ignored the high-frequency terms oscillating at frequency $2\Delta_{2}^{\textnormal{x}}t$. The Langevin-Bloch equations which define the evolution of the spin operators $\sigma_{pq}^{\textnormal{x}}$ can be derived following the methods in \cite{DP3}
We proceed to define the polariton operators as $\Psi _{1,\pm } =\cos \theta^{\textnormal{a}} \hat{E}_{1,\pm } - \sin
\theta^{\textnormal{a}} \sqrt{2\pi n_{z}^{\textnormal{a}}} \sigma^{\textnormal{a}}_{31}$ and $\Psi _{2,\pm } =\cos
\theta^{\textnormal{b}} \hat{E}_{2,\pm } - \sin \theta^{\textnormal{b}} \sqrt{2\pi n_{z}^{\textnormal{b}}}
\sigma^{\textnormal{b}}_{31}$ where $\tan \theta^{\textnormal{a}} = g_{1}^{\textnormal{a}} \sqrt{2 \pi n^{\textnormal{a}}_{z}} / \Omega_{1}$ and $\tan \theta^{\textnormal{b}} = g_{2}^{\textnormal{b}} \sqrt{2 \pi n^{\textnormal{b}}_{z}} / \Omega_{2}$. For simplicity we have assumed that the amplitudes of the counter propagating classical fields are equal, i.e., $\Omega_{i,\pm} \equiv \Omega_{i} (i=1,2)$. In the limit when the excitations are mostly in the spin-wave form, i.e. $\sin \theta^{\textnormal{x}} \simeq 1$, and since $\sigma^{\textnormal{a}}_{31} = -g^{\textnormal{a}}_{1}\hat{E}_{1,\pm }/\Omega _{1}$ and
$\sigma^{\textnormal{b}}_{31} = - g^{\textnormal{b}}_{2}\hat{E}_{2,\pm }/\Omega _{2}$, the polariton operators equal $\Psi_{1,\pm}
= g^{\textnormal{a}}_{1}\sqrt{2\pi n_{z}^{\textnormal{a}}}\hat{E}_{1,\pm }/\Omega _{1}$ and $\Psi_{2,\pm} = g^{\textnormal{b}}_{2}\sqrt{2\pi n_{z}^{\textnormal{b}}}\hat{E}_{2,\pm }/\Omega_{2}$.

Setting $\Psi_{1,2} =(\Psi _{1,2;+}+\Psi _{1,2;-})/2$ and $A_{1,2} = (\Psi _{1,2;+} - \Psi _{1,2;-})/2$  as the  symmetric and antisymmetric combinations of the two polaritons, we derive the equations of motion for the polariton combination $\Psi_{1}, A_{1}$ :
\begin{eqnarray}
\partial_{t}\Psi_{1} + \nu^{(1)} \partial_{z} A_{1} = &-& \sqrt{2\pi}\frac{ (g^{\textnormal{a}}_{1})^{2}}{2\Omega_{1}^{2}} n^{\textnormal{a}}_{z} \partial _{t}\Psi_{1} \notag \\
&-& \mathrm{i} \frac{(g^{\textnormal{a}}_{1})^{2}}{\sqrt{2\pi} \Delta^{\textnormal{a}}_{4}} \left( 2 \Psi_{1}^{\dagger }\Psi_{1}  + A^{\dagger}_{1}A_{1} \right) \Psi_{1} \notag \\
&-& \mathrm{i} \frac{(g^{\textnormal{a}}_{1})^{2} (g^{\textnormal{a}}_{2})^{2} \Omega_{2}^{2}}{\sqrt{2\pi} (g^{\textnormal{b}}_{1})^{2} \Delta_{4}^{\textnormal{a}} \Omega_{1}^{2}} \left( \Psi^{\dagger}_{2}
\Psi_{2} + A^{\dagger}_{2} A_{2} \right) \Psi_{1} \notag \\
&+& \textnormal{noise}, \notag \\
\partial_{t}A_{1} + \nu^{(1)} \partial_{z} \Psi_{1} = &-& \mathrm{i}\sqrt{2\pi} \frac{(g^{\textnormal{a}}_{1})^{2}}{\Delta_{2}^{\textnormal{a}}} n^{\textnormal{a}}_{z} A_{1} - \frac{(g^{\textnormal{a}}_{1})^{2}}{\sqrt{2\pi} \Delta^{\textnormal{a}}_{4}} \Psi^{\dagger }_{1} \Psi_{1} A_{1} \notag \\
&+& \textnormal{noise}
\label{eom}
\end{eqnarray}
A  similar equation holds of the pair $\Psi_{2}, A_{2}$.
$\nu^{(1)}_{g} =\nu^{(1)} \Omega_{1}^{2}/ \pi (g^{a}_{1})^{2} n^{a}_{z}$ and $\nu^{(2)}_{g} =\nu^{(2)} \Omega_{2}^{2}/ \pi (g^{b}_{2})^{2} n^{b}_{z}$ are the corresponding group velocities of the propagating polaritons and $\nu^{(1,2)}$ are the velocities for each quantum field in an empty waveguide. 
The noise terms in Eqs. (\ref{eom}) account for the dissipative processes that take place during the evolution. However for the dark state polaritons under consideration for the case of degenerate lower levels $|1\rangle$ and $|3\rangle$,  and $\Gamma <<  |\Delta^{a}_4|,  |\Delta^{a}_4|$
the losses in the timescales of interest are not significant and thus can be neglected\cite{Chang,Kiffner}. 
Assuming sufficient optical depth of a few thousand and a large ratio between the number of atoms to the number of photon $n^{i}_{z}/\rho_{0.i}\sim10^4$, the antisymmetric combinations $A_{1}$ and $A_{2}$ can be adiabatically eliminated from the equations of motion for the polaritons and moreover, the
nonlinear terms $\Psi _{1}^{\dagger}\Psi _{1}A_{1}$, $\Psi_{2}^{\dagger }\Psi _{2}A_{2}$\ are negligible. Then, Eqs. (\ref{eom})
simplify to
$\partial_{t}\Psi _{1}  =  \frac{1}{2m_{1}}\partial_{z}^{2}\Psi _{1} + U_{1}
\Psi_{1}^{\dagger }\Psi _{1} \Psi_{1} + V_{1}\Psi_{2}^{\dagger }\Psi _{2}\Psi _{1}$ and $
\partial _{t}\Psi _{2}  =  \frac{1}{2m_{1}}\partial_{z}^{2}\Psi _{2} + U_{2}
\Psi_{2}^{\dagger }\Psi _{2} \Psi_{2} + V_{2}\Psi_{1}^{\dagger }\Psi _{1}\Psi _{2}$
These are the equation of motion of a two-component Luttinger liquid of polaritons with the corresponding Hamiltonian:
\begin{eqnarray}
H & = & \hbar \int dz \left\{ \sum_{i}\left[ \frac{1}{2m_{i}}\partial _{z}\Psi_{i}^{\dagger }( z)
 \partial_{z}\Psi _{i}( z) +\frac{U_{i}}{2} \rho _{i}^{2}(z) \right]  \right. \notag \\
& & \left. +  V_{12} \rho _{1}(z)\rho_{2}(z) \right\}
\end{eqnarray}
The interaction parameters relating to the  effective kinetic energy, intra and inter species repulsion can be tuned by controlling the external control fields and the various detunings. $m_{i}$ is effective mass for the $i$th polariton with
 $\frac{1}{m_{1}}=-\frac{4\Delta^{a}_{2} \nu^{(1)}_{g}}{\Gamma^{a}_{1D} n^{a}_{z}}$ and
  $\frac{1}{m_{2}}=-\frac{4\Delta^{b}_{2} \nu^{(2)}_{g}}{\Gamma^{b}_{1D} n^{b}_{z}}$ with $\Gamma^{a,b}_{1D}$
 the spontaneous emission rate of a single atom in the waveguide modes. The intra  repulsion are given by $U_{1}=\frac{\Gamma_{1D} \nu^{(1)}_{g} } {2\Delta^{a}_{4}} $ $U_{2}=\frac{\Gamma_{1D} \nu^{(2)}_{g} } {2\Delta^{b}_{4}} $
%
 and
  the inter-species by $V_{12}=V_{1}+V_{2}$ and $V_{1} =\frac{\pi \left( g^{a}_{1}\right) ^{2}\left(g^{a}_{2}\right) ^{2}\nu_{g}^{(1)}}{ \left( g^{b}_{2}\right) ^{2} \Delta _{4}^{a}\nu^{(1)}}$ with  $V_{2} =  \frac{\pi (g^{b}_{2}) ^{2}\left( g^{b}_{1)}\right) ^{2}\nu_{g}^{(2)}}{ \left( g^{a}_{1}\right) ^{2} \Delta _{4}^{b}\nu^{(2)}}$.

Let us now define  the effective parameters $u_{i}=\sqrt{\rho _{0,i}U_{i}/m_{i}}$ and $K_{i}=\sqrt{\pi ^{2}\rho _{0,i}/\left( m_{i}U_{i}\right) }$ with $\rho _{0,i}$\  equal to $n_{ph,i}$, the photon number of $i$th\ quantum field in our scheme. It is known from the works by Girardeau and others,   that when  $u_{1}=u_{2}=u$ and $K_{1}=K_{2}=K$ the above LL Hamiltonian Eq. 1 can be transformed into new one comprised separately of two parts, the charge part $H_{c}$ and the spin part $H_{s}$ as $H =H_{c}+H_{s}$ \cite{Girardeau}. This allows for the separation of a single excitation into two separate ones each comprised of spins or of charge/density. These can propagate through the liquid with different velocities given by $u_{c,s} =u\sqrt{1\pm \frac{\left( V_{1}+V_{2}\right) K}{\pi u}}$.
In our case, the charge(spin) density corresponds to the sum(difference)  of the corresponding polaritons densities
$n_{c,s}=n_{1} \pm n_{2}$ with $n_{i}=\langle \Psi^{\dagger}_{i} \Psi_{i} \rangle $ and $\Psi_{i} =(\Psi _{i;+}+\Psi _{i;-})/2$ the symmetric combinations of the two counter-propagating dark state polaritons generated by each set of atom-field interaction. 

We note here that in our case, Eq. 1 describes a non-equilibrium situation where we follow the dynamic evolution of two polariton pulses. Here the effective masses $m_{1,2}$ are kept constant in time while the effective inter and intra species repulsions $U_{1,2}$ and $V_{12}$ are increased to reach the necessary regime.
This is possible by keeping for example the  corresponding detunings  $\Delta^{a}_{2}$,  $\Delta^{b}_{2}$ constant, while shifting the level 4 responsible for the nonlinear shifts by changing the detunings $\Delta^{a}_{4}$,$\Delta^{b}_{4}$. Labeling $\gamma_{1,2}$ the corresponding ratio of the interaction to the kinetic energies for each polariton species, at the time of the pulses  entering the fiber, conditions are such that $\gamma_{1,2}<1$. At this phase interactions are weak and the photons expand freely due to dispersion. By shifting in the 4th-levels, the system can be driven into the strongly interacting regime $\gamma_{1,2}>1$ necessary for spin charge separation to occur. 
\footnote{We note here that throughout the process we should keep $\Delta _{2}^{a}$ and$\ \Delta _{2}^{b}$\ are negative (positive) and $\Delta_{4}^{a}$\ and $\Delta _{4}^{b}$\  positive (negative)  in order to keep the corresponding interaction terms to be always positive.}.
\begin{figure}
\centering
\includegraphics[scale=0.5]{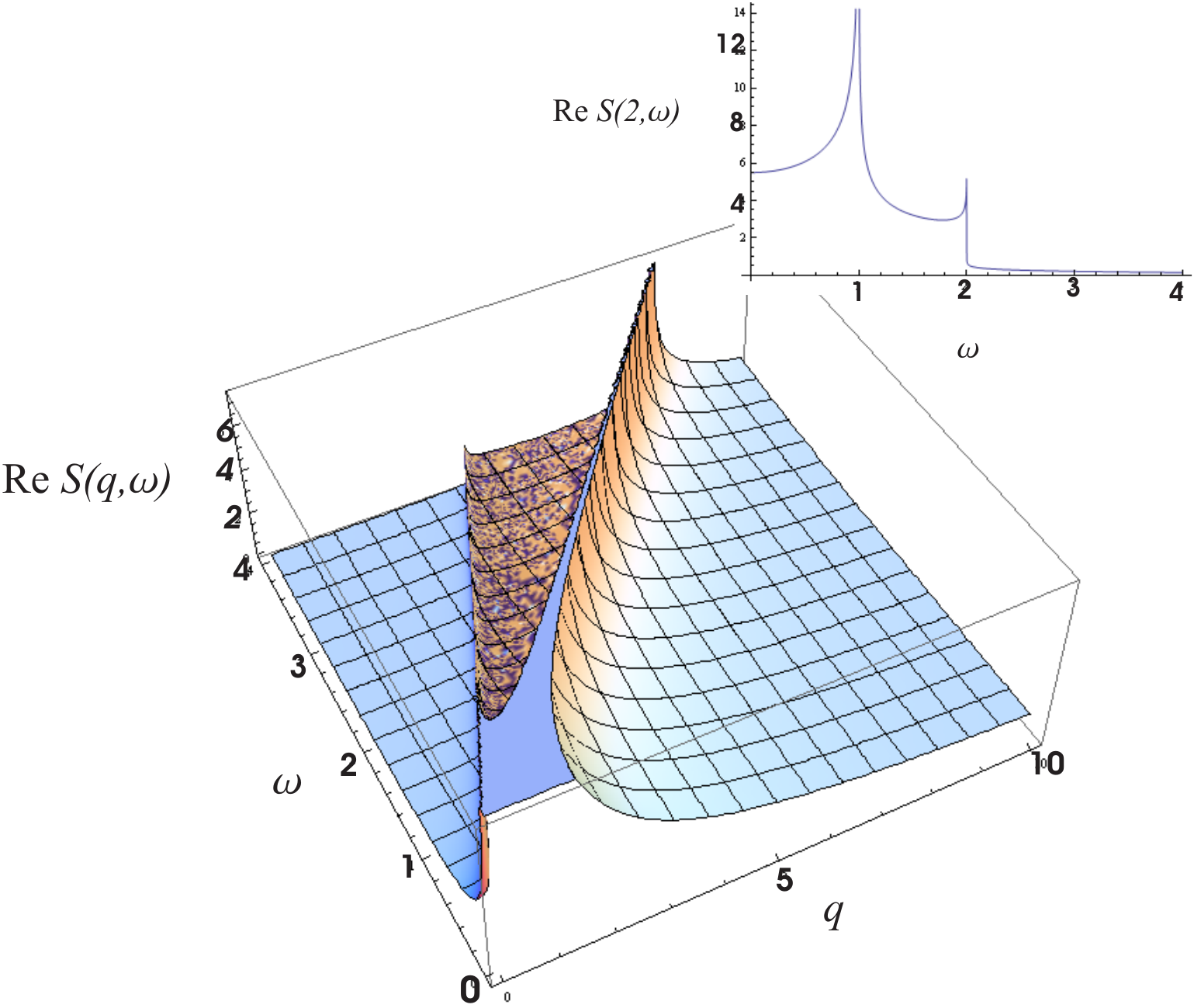}
\caption{The single particle spectral function for our photonic system. The effective spin and charge velocities are $u_{s}=0.5$ and $u_{c}=1$ and can be achieved for optical depths around OD=4000, with 10 photons in each pulse initially, and single atom cooperativity for each atomic species of 0.2. Subsequent releasing of one of the trapped polaritons through its  coherent mapping to a propagating light pulse, allows for the efficient measurement of the spectral function.This is done though standard quantum optical measurement techniques measuring correlations in the intensities of the output fields
 and by analyzing the optical spectrum. In the inset we plot a
2D cut for the value of q=2. The units of  $\omega$ and q are in $\frac{\pi }{z_{0}}\times 2\sqrt{\frac{2}{5}}u$ and $\pi/z_{0}$ respectively.}
\label{fig2}
\end{figure}

We will now analyze in detail the steps characterizing the preparation, release, and the measurement of the separation, estimating the quantum optical parameters required to reach the LL regime. First the fiber is injected with the two quantum coherent pulses  $E_{1,+}, E_{2,+}$ from the left and the adiabatic switching off the two classical fields $\Omega _{1,+}$ and $\Omega _{2,+}$ occurs. The polaritons $\Psi_{1}$ and $\Psi_{2}$ are formed and get trapped. Then,  all four classical fields $\Omega _{1,\pm}$ and $\Omega _{2,\pm}$ are slowly
switched on so that an effective Bragg grating is formed. The latter forces the polaritons -- which now have acquired a small photonic component as well to evolve under the LL model. The interaction parameters (couplings and detunings)
characterizing the intra- and inter- species repulsion can be tuned  in time in order to reach the necessary regime for density separation. Once this is achieved one field, say  say $\Omega_{1+}$, from the pair of control fields  that trap polariton type $\Psi_{1}$ is  slowly turned off. This will release the corresponding quasiparticle and allow it to propagate towards the exit of the fiber. As all correlations established in the previous step- the  evolution under the LL Hamiltonian Eq. 1-are retained, this wavepacket containing mostly of light now, comprises of two separated  parts propagating with different velocities $u_{s}$ and $u_{c}$ towards the end of the fiber.
There are two ways to observe the separation in this system. The first is by measuring the time evolution
of a single excitation as usually predicted for  the case of cold atoms proposals as well\cite{Recati,Kollath, Kleine}. In our case the charge(spin) density waves, after the release of the polaritons, will transfer to the sum(difference)  of the corresponding time dependent photon intensities for each propagating field
$n_{c,s}=n_{1} \pm n_{2}$ with $n_{i}=\langle E^{\dagger}_{i,+} E_{i;+} \rangle $. Using standard optical measurement techniques one could dynamically probe the time evolution of the corresponding effective 
charge and spin densities maxima and from that infer the corresponding velocities.
The second way to measure the effect is through directly probing the correlations established during the evolution with the LL Hamiltonian by measuring the corresponding {\it single} particle spectral function as usually proposed in
condensed matter experiments
$S(q,\omega)$ \cite{Girardeau,metallic, organic,SCC3,SCC4}. In our case, as the quasiparticle after the coherent release are propagating photo pulses, this is directly accessible by measuring the first order correlation function ( which corresponds to analyzing the spectrum) of one of the initial fields, say $E_{1}$ for a specific quasimomenta q. This should exhibit two peaks as shown in the inset of Fig. 2 centered around  $u_{s}q$ and $u_{c}q$ with $q$ the corresponding quasi-momentum of the propagating excitation. In our case q is inversely proportional to the initial extend of the pulse $z_{0}$. For clear distiction between the two spin and charge peaks we should tune our optical detectors around  $q=2\pi/z_{0}$ . In Fig. 2 we plot the spectral function as analytically derived for a two component bosonic system\cite{Iucci}, for intra and inter species values corresponding in our system to $u=1$ and $k=1$. Here $U_{1}/V_{12}=U_{2}/V_{12}=0.6$ which in turn give a ratio between the effective charge and spin velocities of $u_{c}/u_{s}=2$. 
The corresponding necessary values of the parameters  in our system are optical depths OD=4000  and single atom cooperativity of 0.2 and roughly 10 photons initially each pulse\footnote{single-atom cooperativity $\eta^{a,b} = \Gamma_{1D}/ \Gamma$, which describes the ratio of emission into the
waveguide to total emission, and an optical depth $OD^{a,b} = \eta n^{a,b}_{z}z^{a,b}_{0}$. Here  $\Gamma^{a,b}$ is the total spontaneous emission rate of the excited states of the atoms (which for simplicity we assume the same), $n^{a,b}_{z}$ is the density of atoms coupled to the waveguide, and $z^{a,b}_{0}$ is the characteristic length of the pulses initially.}. We note here in calculating the latter we took into account that both the linear and the nonlinear loss mechanisms will define a maximum evolution time $t_{\textnormal{max}}$ for the second phase of the process. This in turn will give the following condition on
the achievable ratio of interaction to kinetic energies \cite{Chang,Kiffner}
 \begin{equation}
\gamma^{1,2}_{\textnormal{max}} \sim \min \big(\gamma_{0} \exp(\frac{\beta |\Delta^{a,b}_{2}|}{\Gamma},\eta \beta \frac{\Gamma}{|\Delta^{a,b}_{2}|} \frac{OD^{a,b}}{N^{1,2}_{\textnormal{ph}}}),
\end{equation}
Optimizing over $\Delta^{a,b}_{2}$ will give the numbers mentioned above. We add here that in order to omit the noise terms and also be able to neglect the higher order derivatives so that the system is within good approximation a two component Luttinger liquid as described in Eq. 1, the following conditions need to be satisfied.  The Stark shift of levels $|2\rangle$ for each atomic species should stay within the EIT transparency window and  a bound on spin-wave excitations wavevector should
be in place. These conditions translate to 
$ n_{1}/n^{a}_{z} \ll |\Delta^{a}_{4}|/|\Gamma-2i\Delta^{a}_{2}|$,$
 n_{2}/n^{b}_{z} \ll |\Delta^{b}_{4}|/|\Gamma-2i\Delta^{b}_{2}| 
$  and
$n_{1}/n^{a}_{z} \ll \Gamma_{1D}/|2\Delta^{a}_{2}+i\Gamma|$,$
n_{2}/n^{b}_{z} \ll \Gamma/|2\Delta^{a}_{2}+i\Gamma|
$ which is satisfied in our case as the numbers of atoms can be two or three order of magnitude larger than the
number of photons.
We note that in this case the ratio between kinetic and repulsion energies for
each species can reach the value of 40 which for the single component case was shown to
lead to a Tonks gas of photons \cite{Chang}. For spin-charge separation, the required repulsions could be
of a smaller value, thus relaxing the overall quantum optical requirements but the two peaks in the spectral function
will be less pronounced.

In conclusion we have shown   that stationary light-matter excitations generated inside a hollow one-dimensional  waveguide filled with  atoms, can be made to simulate a photonic two-component Lieb Liniger model. Moreover  the relevant interactions exhibit the necessary tunability  both to generate and efficiently observe spin charge separation using standard quantum optical methods.

We would like to acknowledge financial support by the National Research Foundation \&
Ministry of Education, Singapore.




\begin{thebibliography}{99}

\bibitem{Girardeau} {\small M. Girardeau, \emph{J. Math. Phys.} (N.Y.) \textbf{1},
516 (1960); B. Paredes, \emph{et al.}, \emph{Phys. Rev. A} \textbf{66},
033609 (2002); M. Girardeau, \emph{Phys. Rev.} \textbf{139}, 500 (1965);
N.M. Bogoliubov, V.E. Korepiin, and A.G. Izergin, in \emph{Quantum Inverse
Scattering Method and Correlation Functions} (Cambridge University Press,
Cambridge, UK, 1993); T. Giamarchi  \emph{Quantum Physics in One Dimension} (Oxford University Press,
Oxford,UK, 2003).
}


\bibitem{metallic} {\small P. Segovia, D.Purdie, M. Hengsberger, and Y. Baer,
\emph{Nature} (London) \textbf{402}, 504 (1999).}

\bibitem{organic} {\small T. Lorenz, \emph{et al.}, \emph{Nature} (London)
\textbf{418}, 614 (2002).}



\bibitem{SCC3} {\small C. Kim \emph{et al.}, \emph{Phys. Rev. Lett.} \textbf{%
77}, 4054 (1996).}

\bibitem{SCC4} {\small O.M. Auslaender \emph{et al.}, \emph{Science} \textbf{%
308}, 88 (2005); Y. Jompol \emph{et al.}, \emph{Science} \textbf{%
325}, 597 (2009). }

\bibitem{Bloch} {\small M. Greiner, O. Mandel, T.Esslinger, Th.W. H\"{a}nsch,
and I. Bloch, \emph{Nature} (London) \textbf{415}, 39 (2002).}

\bibitem{Jaksch} {\small D. Jaksch, C. Bruder, J.I. Cirac, C.W. Gardiner, and
P. Zoller, \emph{Phys. Rev. Lett.} \textbf{81}, 3108 (1998).}

\bibitem{ions}
D. Porras and I. Cirac, Phys. Rev. Lett. {\bf 92}, 207901 (2004);
A. Friedenaue  \emph{et al.}. Nature Physics {\bf 4}, 757 - 761 (2008).

\bibitem{angelakis07} {\small D. G. Angelakis, M. F. Santos, and S. Bose,
Phys. Rev. A \textbf{76}, 031805(R) (2007).}

\bibitem{hartmann06} {\small M. J. Hartmann, F. G. S. L. Brand\~ao, and M.
B. Plenio, Nature Phys. \textbf{2}, 849 (2006).}

\bibitem{greentree06} {\small A. D. Greentree, C. Tahan, J. H. Cole, and L.
C. L. Hollenberg, Nature Phys. \textbf{2}, 856 (2006).}

\bibitem{Chang} {\small D. E. Chang, V. Gritsev, G. Morigi, V. Vuleti\'{c},
M. D. Lukin, and E. A. Demler, \emph{Nature Physics} \textbf{4}, 884 (2008).}

\bibitem{Aichhorn} {\small M. Aichhorn \emph{et al.}, \emph{Physical Review
Letters} \textbf{100}, 216401 (2008). }

\bibitem{coherent control of photon emission} {\small Dario Gerace \emph{et
al.}, \emph{Nature Physics} \textbf{5}, 281 (2009); I. Carusotto \emph{et al.%
}, \emph{Physical Review Letters} \textbf{103}, 033601 (2009). }



\bibitem{Recati} {\small A. Recati, P.O. Fedichev, W. Zwerger, and P. Zoller,
\emph{Phys. Rev. Lett.} \textbf{90}, 020401 (2003).}

\bibitem{Paredes03} {\small B. Paredes and J.I. Cirac, \emph{Phys. Rev. Lett.}
\textbf{90}, 150402 (2003).}

\bibitem{Paredes04} {\small B. Paredes, \emph{et al.}, \emph{Nature} \textbf{429},
277-281 (2004).}

\bibitem{Kecke} {\small L. Kecke, H. Grabert, and W. Hausler, \emph{Phys.
Rev. Lett.} \textbf{94}, 176802 (2005).}

\bibitem{Kollath} {\small C. Kollath, U. Schollw\"{o}ck, and W. Zwerger, \emph{%
Phys. Rev. Lett.} \textbf{95}, 176401 (2005).}

\bibitem{Kleine} {\small A. Kleine, \emph{et al.}, \emph{Phys. Rev. A} \textbf{%
77}, 013607 (2008).}

\bibitem{DP1} {\small M. Fleischhauer and M.D. Lukin. \emph{Phys.
Rev. Lett.} \textbf{84}, 5094–5097 (2000).}

\bibitem{DP2} {\small M. Bajcsy, A.S. Zibrov, and M.D. Lukin,. \emph{Nature}
\textbf{426}, 638–641 (2003).}

\bibitem{DP3} {\small M. Bajcsy, \emph{et al.}, \emph{Phys. Rev.
Lett.} \textbf{102}, 203902 (2009).}

\bibitem{Kiffner} {\small M. Kiffner and M. Hartmann,  \emph{Phys. Rev. A} 
\textbf{81}, 021806 (2010).}

\bibitem{Iucci} {\small A. Iucci, G.A. Fiete, and T. Giamarchi,
\emph {Phys. Rev. B} \textbf{75}, 205116 (2007).}


\end{thebibliography}
\end{document}